\documentstyle[12pt,epsf]{article}

\def\ptslash{{p_{t}}{\kern -1.3ex\hbox{/}}}

\begin{document}
\baselineskip=18pt

\def\gsim{\lower0.5ex\hbox{$\:\buildrel >\over\sim\:$}}
\def\lsim{\lower0.5ex\hbox{$\:\buildrel <\over\sim\:$}}
\def \rp{{R\hspace{-0.22cm}/}_P}
\def \lp{{L\!\!\!/}}
\def \n{\noindent}

\begin{flushright}
%{AMES-HET-04-??}\\
{BNL-HET-04/10}\\
\end{flushright}

\begin{center}
{\large\bf
Seesaw induced electroweak scale, the hierarchy problem and 
sub-eV neutrino masses}
\end{center}

\vspace*{0.5 in}

\renewcommand{\thefootnote}{\fnsymbol{footnote}}
\centerline{David Atwood\footnote{Electronic address: atwood@iastate.edu}}
\centerline{\it Dept. of Physics and Astronomy, Iowa State University, Ames,IA 50011}
\vspace{0.1 in}
\centerline{Shaouly Bar-Shalom\footnote{Electronic address:shaouly@physics.technion.ac.il}}
\centerline{\it Physics Department, Technion-Institute of Technology, Haifa 32000, Israel}
\vspace{0.1 in}
\centerline{Amarjit Soni\footnote{Electronic address: soni@bnl.gov}}
\centerline{\it High Energy Theory Group, Brookhaven National Laboratory, Upton, NY
11973}
\vspace{0.1 in}
\renewcommand{\thefootnote}{\arabic{footnote}}
\setcounter{footnote}{0}
\vspace*{0.5 in}

\begin{center}
{\bf Abstract}
\end{center}

We describe a model for the scalar sector where all interactions occur
either at an ultra-high scale $\Lambda_U \sim 10^{16}-10^{19}$ GeV or 
at an intermediate scale $\Lambda_I=10^{9}-10^{11}$ GeV. 
The interaction of physics on these two scales 
results in an SU(2) Higgs condensate at the electroweak (EW) scale, 
$\Lambda_{EW}$, through a seesaw-like Higgs mechanism, 
$\Lambda_{EW} \sim \Lambda_I^2/\Lambda_U$, while the breaking of 
the SM SU(2)$\times$U(1) gauge symmetry 
occurs at the intermediate scale $\Lambda_I$.
The EW scale is, therefore, not fundamental but is
naturally generated in terms of ultra-high energy phenomena and so
the hierarchy problem is alleviated. We show that the class of such
``seesaw Higgs'' models predict the existence of 
sub-eV neutrino masses which are 
generated through
a ``two-step'' seesaw mechanism in terms of the same 
two ultra-high scales: 
$m_\nu \sim \Lambda_I^4/\Lambda_U^3 \sim \Lambda_{EW}^2/\Lambda_U $.
The neutrinos can be either 
Dirac or Majorana, depending 
on the structure of the scalar potential.
We also show that our seesaw Higgs model
can be naturally embedded in theories with {\it tiny} extra dimensions
of size $R \sim \Lambda_U^{-1} \sim 10^{-16}$ fm, 
where the seesaw induced EW scale 
arises from a violation of a symmetry at a distant brane; in particular, 
in the scenario presented there are 7 {\it tiny} extra dimensions.

\newpage

\section{Introduction}\label{intro}

A long standing problem in modern particle physics is the 
apparent enormous hierarchies of energy/mass scales observed 
in nature. Disregarding the ``small'' hierarchies in the masses of the 
known charged matter particles, there seems to be two much larger hierarchies:
the first is the hierarchy between the fundamental grand unified scale 
$\Lambda_U \sim {\cal O}(10^{16})$ GeV [or Planck scale 
$\Lambda_U \sim {\cal O}(10^{19})$ GeV], and the EW scale, 
$\Lambda_{EW} \sim {\cal O}(100)$ GeV, and the second is the   
hierarchy between the EW scale and the neutrino mass scale 
$m_\nu \sim {\cal O}(10^{-3})$ eV. 
Denoting the former as the Upper Hierarchy (UH) and 
the latter as the Lower Hierarchy (LH), it is interesting to note
that these two hierarchies (the UH and LH) 
span over roughly the same energy scales, i.e., over about
14 orders of magnitude. This hierarchical structure of scales 
raises some unavoidable questions: What is the physics behind 
EW symmetry breaking (EWSB)? Is it the same physics that 
underlies the neutrino mass scale? Does the scale of the new physics 
lie close to the EW scale or close to the GUT or Planck scale?
and finally, is there a theory that can provide 
a natural explanation for the simultaneous presence of 
$\Lambda_U,~\Lambda_{EW}$ and $m_\nu$? These questions 
have fueled a lot of activity in the past 30 years in the search for new 
physics beyond the Standard Model (SM).

An example of a successful mechanism for explaining the LH 
(i.e., $m_\nu / \Lambda_{EW}$) was suggested a long time ago
\cite{seesaworig} - the so called seesaw mechanism in which 
the new physics is assumed to lie at the ultra-high fundamental scale 
$\Lambda_U$. Given the EW scale, 
this mechanism utilizes the UH (i.e., $\Lambda_{EW}/\Lambda_U$) 
to solve the LH, since the seesaw generated neutrino mass scale 
is $m_\nu \sim (\Lambda_{EW}/\Lambda_U) \times \Lambda_{EW}$.

The UH, when viewed within the SM framework, is often called the 
Hierarchy Problem (HP) of the SM, which is intimately related to 
the SM Higgs sector responsible for the generation of the 
EWSB scale, $v_{EW} \sim \Lambda_{EW}$, through the Higgs mechanism. 
The HP of the SM raises a technical difficulty known as the naturalness 
(or fine tuning) problem. Simply put,       
the presence of a fundamental EW scale seems to be unnatural,
since if the only physics which exists up to some ultra-high grand
unified scale $\Lambda_U$ is the SM, then 
it is hard to see why the Higgs boson does not
receive large corrections to its mass. In other words, 
there is a problem of stabilizing the ${\cal O}(\Lambda_{EW})$ 
mass scale of the Higgs against radiative corrections without 
an extreme fine tuning (to one part in $\Lambda_{EW}^2/\Lambda_U^2$) between 
the bare Higgs mass squared and the ${\cal O}(\Lambda_U^2)$ corrections, $\Lambda_U$ being 
the grand unified or Planck scale.  

This fine tuning problem of the Higgs mass term has been
addressed by numerous models for beyond the SM physics. 
For instance, in 
weak scale Supersymmetry (SUSY) models, in which the SUSY breaking 
scale $M_{SUSY}$ is close to the EW scale, such large 
corrections to the Higgs mass 
are suppressed, due to a symmetry between fermions and bosons giving rise 
to natural
cancellations between graphs involving particles and superparticles 
\cite{mhsusy}.
Likewise, in Technicolor inspired models, new  
gauge 
interactions dynamically generate a new scale $M_{dyn}$ from which the 
EWSB is generated. In this scenario
the light Higgs is usually viewed
as a pseudo-Goldstone boson associated with the breaking of
the new strong gauge force and the HP is alleviated 
if $M_{dyn}$ is close to $\Lambda_{EW}$.
In this case, somewhat like in the SUSY case, 
it has been recently argued that one may be able to explain away
the little fine tuning left (i.e., between $M_{dyn}$ and $\Lambda_{EW}$)
by ``Little Higgs'' models \cite{LHM}, through  
loops involving additional particles with TeV scale masses 
that cancel the SM loop graphs. Contrary to SUSY models, 
in Little Higgs models these cancellations occur 
between particles of the same statistics. 

Another interesting approach for addressing the fine tuning 
problem of the SM was suggested in \cite{bardeen}. According to 
\cite{bardeen}, the scale invariance of the classical level SM
can be used to remove the explicit quadratic divergences in the 
higher order corrections to the Higgs 
mass order by order in perturbation theory, 
without invoking any fine tuning. In this case,
both the bare Higgs mass 
and the Higgs mass counter term can be chosen at the EW scale and so 
the conditions on the Higgs mass are natural.
  
However, SUSY and Little Higgs models as well as the scale invariance 
theorem do not provide a solution to the 
question of the origin of scales, i.e., why do we observe 
such a large hierarchy between the 
fundamental GUT or Planck scale $\Lambda_U$ and the EW scale $\Lambda_{EW}$ 
(recall that $M_{SUSY} \sim \Lambda_{EW}$ in the SUSY case 
and $M_{dyn} \sim \Lambda_{EW}$ in the Technicolor-like 
models case)? Of course, these models do give 
a successful resolution to the technical difficulty of explaining 
the simultaneous presence of the two disparate scales 
$\Lambda_U$ and $\Lambda_{EW}$, thus solving the naturalness or fine tuning 
problem.     
In contrast, in extra dimensional scenarios such questions are muted
since these theories contain only one fundamental scale and so the hierarchy 
between the Planck and the EW scales is simply absent. 
For example,  
in large extra dimensional models \cite{LED} 
the EWSB  scale $\Lambda_{EW}$ is considered to be the only 
fundamental scale and the observed enormity 
of the Planck scale is a consequence of the large extra dimensions 
through which gravity propagates. 
Alternatively, in models where the four-dimensional metric is 
multiplied by a ``warp'' exponential (rapidly 
changing) factor of one tiny extra dimension \cite{RS99}, 
the Planck scale is viewed as the fundamental scale 
and the EW mass scale is generated due 
to this exponential suppression factor.  
     
In this paper we propose an alternative approach, 
where the only fundamental scale is the GUT or Planck 
scale $\Lambda_U$, 
while the EW and neutrino mass scales both arise due to interactions between
the fundamental scale $\Lambda_U$ and a new intermediate ultra-high scale
$\Lambda_I \sim 10^{9}-10^{11}$ GeV, i.e., 
$\Lambda_{EW}<<\Lambda_I<<\Lambda_U$. The intermediate scale   
is viewed as the
scale of breaking of the unification group which underlies the physics at 
the scale $\Lambda_U$ (see e.g., \cite{mohapatra}).   
In particular, this class of models (see also \cite{calmet}) 
are based on the idea that 
EWSB occurs 
at $\Lambda_I$ whereas the smallness of the EW mass scale is 
a consequence of a $\Lambda_U-\Lambda_I$ seesaw-like Higgs 
mechanism in the scalar 
potential, giving $\Lambda_{EW} \sim \Lambda_I^2/\Lambda_U$, from which 
the masses of all known particles are generated. 
In addition, our model naturally accounts for the existence of sub-eV 
neutrino masses by means of a ``two-step'' or ``double'' 
seesaw mechanism using the 
same two ultra-high scales $\Lambda_U$ and $\Lambda_I$: the first $\Lambda_U-\Lambda_I$ seesaw generates the
EW scale $\Lambda_{EW} \sim \Lambda_I^2/\Lambda_U$ and then a second $\Lambda_U-\Lambda_{EW}$ seesaw gives 
rise to the sub-eV neutrino mass scale 
$m_\nu \sim \Lambda_{EW}^2/\Lambda_U \sim \Lambda_I^4/\Lambda_U^3$.
Thus, in this type of ``seesaw Higgs'' models 
the large desert(gap) 
between the fundamental scale $\Lambda_U \sim 10^{16}$ or $10^{19}$ GeV 
and the EW scale $\Lambda_{EW}$ (the UH) and between the EW scale 
and the neutrino mass scale (the LH) 
are both naturally explained 
by introducing an intermediate ultra high scale, 
$\Lambda_I \sim 10^9 - 10^{11}$ GeV, in which new interactions 
are manifested {\it only in the scalar sector}.

We will present one possible minimal variation of seesaw Higgs models
which we will name the ``light seesaw'' model. 
In our light seesaw model
the scalar spectrum consists of several physical scalars with masses
of order $\Lambda_U$ and one SM-like Higgs with a mass 
of ${\cal O}(\Lambda_{EW})$.
The light seesaw model will be minimally constructed in the sense that at 
energy scales of ${\cal O}(\Lambda_{I})$ and below, i.e., 
much smaller than the fundamental scale 
$\Lambda_U$, it contains the SM 
gauge symmetries and fields apart from 
the addition of new (superheavy) right handed neutrino fields.  
Thus, at energies of ${\cal O}(\Lambda_{EW})$, the light seesaw model
which may be considered as a minimal extension of the SM without a HP,
will have the same phenomenological signatures as the ``one-Higgs'' SM.
This implies that, at the LHC, only one SM-like Higgs state will be observed 
in contrast to e.g. the expectations from SUSY models in which 
several Higgs states should be detected.
In addition, due to the presence of the superheavy right-handed neutrinos, 
our light seesaw model will contain sub-eV neutrino masses in
accord with recent measurements \cite{neutrino1}.
 
We also show that seesaw Higgs models can emanate from theories with 
extra spatial dimensions. However, contrary to 
large extra dimensions models in which the $4+n$ 
fundamental Planck 
scale ($n$ is the 
number of extra dimensions)
is taken to be of ${\cal O}(\Lambda_{EW})$, 
the seesaw Higgs mechanism requires
the fundamental scale to be of ${\cal O}(\Lambda_U)$, which results 
in {\it tiny} compact extra dimensions of size $R \sim \Lambda_U^{-1}$.
This somewhat resembles the ``warp'' extra dimension model of \cite{RS99}
which also requires a tiny extra dimension with a 
fundamental scale of order of the 
Planck scale.   
We present a ``tiny extra dimensions'' scenario that can 
naturally explain the existence of an intermediate 
scale $\Lambda_I \sim 10^9$ GeV, required for triggering 
the seesaw Higgs mechanism.
In particular, in this scenario the intermediate scale $\Lambda_I$  
is generated due to violation of some symmetry at a distant brane. 

We wish to emphasize
that the light seesaw Higgs model presented in this paper
does not attempt to represent the complete theory but,
only to parametrize its low energy dynamics and
to provide a schematic model that holds the key ingredients for 
model building of a more complete underlying seesaw Higgs model. 

Finally, we note that a scalar model that falls into the category of 
light seesaw models was also
proposed by Calmet in \cite{calmet}. The Calmet model is 
re-examined in this paper since it demonstrates the seesaw Higgs 
mechanism with only SU(2) scalar doublets, whereas in our light seesaw model 
the seesaw Higgs mechanism is based on interactions between SU(2) doublets 
and singlets. Moreover, the issue of neutrino masses was not addressed in 
\cite{calmet}. It turns out that the Calmet model is 
complimentary to ours since 
it predicts Dirac neutrinos whereas our model gives rise to 
Majorana neutrinos.  
We also note that the idea of a seesaw mechanism which originates from 
a scalar sector was applied before in the so called ``top quark 
seesaw models'', in which the EWSB is triggered by a condensation 
which appears due to strong topcolor gauge interactions. In these models, 
the right size of the top quark mass is ensured by a seesaw structure in 
the masses of the composite states \cite{topseesaw}.
A seesaw mechanism in the scalar sector was also applied to SUSY models 
as a possible solution to the $\mu$-problem \cite{muprob}.    
       
The paper is organized as follows: in section II we describe the 
seesaw Higgs mechanism as it is manifested in the light seesaw model. 
In section III we
discuss the generation of neutrino masses in seesaw Higgs models,
in section IV we discuss the possibility of embedding 
the light seesaw model in theories
with extra dimensions and in section V we summarize.

\section{The seesaw Higgs mechanism}\label{seesaw}

As mentioned above, in the seesaw Higgs
models presented below, the EW scale $\Lambda_{EW}$ 
is not fundamental since its existence is triggered
by physics at much higher energy scales (and therefore more
fundamental). 
In particular, $\Lambda_{EW}$ is generated 
by the large splitting between the fundamental scale $\Lambda_U$ 
and an intermediate scale $\Lambda_I$, such that:

\begin{eqnarray}
\epsilon \equiv \Lambda_I/\Lambda_U \sim 10^{-7} ~{\rm or}~ 10^{-8.5}
~,
\end{eqnarray}

\noindent depending 
on whether the fundamental scale $\Lambda_U$ is taken to be 
around the GUT scale 
[$\Lambda_U \sim {\cal O}(10^{16})$ GeV] or around the Planck scale 
[$\Lambda_U \sim {\cal O}(10^{19})$ GeV], respectively.

Let us schematically define the seesaw Higgs total Lagrangian as:

\begin{eqnarray}
{\cal L}={\cal L}_{SM}(f,G)+ {\cal L}_S(\Phi,S) + 
{\cal L}_{Y}(\Phi,f) + {\cal L}_\nu(\Phi,S,\nu_R) \label{totlag} ~,
\end{eqnarray}   

\noindent where ${\cal L}_{SM}(f,G)$ is the usual SM's fermions and 
gauge-bosons kinetic terms, ${\cal L}_S(\Phi,S)$
contains the seesaw-like scalar potential as well as the 
kinetic terms for the SU(2) 
doublet $\Phi$ and the scalar singlets $S$, 
${\cal L}_{Y}(\Phi,f)$ is the SM Yukawa 
terms and 
${\cal L}_\nu(\Phi,S,\nu_R)$ contains both Dirac and Majorana-like 
interactions between the scalars and the left and right-handed neutrinos
($\nu_R$).
In what follows we will consider two types of 
${\cal L}_S(\Phi,S)$ that can give rise to a seesaw Higgs
mechanism; one that leads to our light seesaw model and another that 
leads to the seesaw Higgs model discussed in \cite{calmet} - the 
``Calmet model''.

\subsection{The light seesaw model}\label{LHSM}

Consider the following scalar Lagrangian:

\begin{eqnarray}
{\cal L}_S(\Phi,\varphi,\chi)= |D_\mu \Phi|^2 + |\partial_\mu \varphi|^2 +
|\partial_\mu \chi|^2 - V(\Phi,\varphi,\chi) \label{lscal} ~, 
\end{eqnarray}

\noindent where $\Phi$ is an SU(2) doublet and $\varphi,~\chi$ are 
``sterile'' SU(2) singlets
which do not interact with the SM fields.
A scalar potential containing the two singlets 
and one doublet in (\ref{lscal}) and subject to 
a seesaw Higgs mechanism can be constructed as follows:

\begin{eqnarray}
V(\Phi,\varphi,\chi) = &&\lambda_1 \left( |\Phi|^2 - |\chi|^2 \right)^2 +
\lambda_2 \left( |\varphi|^2 - \Lambda_U^2 \right)^2 \nonumber \\ 
&&+  ~ \lambda_3 \left( {\rm Re}(\varphi^\dagger \chi) - \Lambda_I^2 \cos\xi 
\right)^2 
+ \lambda_4 \left({\rm Im}(\varphi^\dagger \chi) - \Lambda_I^2 \sin\xi 
\right)^2 \label{hpotlittle} ~,   
\end{eqnarray}

\noindent where, we have assumed a massless SU(2) doublet field
$\Phi$, one massless singlet field $\chi$ and one superheavy
singlet field $\varphi$.\footnote{The massless SU(2) doublet 
$\Phi$ and singlet $\chi$
may be Goldstone bosons related to a spontaneously broken global symmetry 
at the high scale $\Lambda_U$.}
Also, all
$\lambda_i$ are positive real constants, naturally of ${\cal O}(1)$,
and $\xi$ is a possible
relative phase between the VEVs of $\varphi$ and $\chi$ which may give rise
to spontaneous CP-violating effects, e.g., in the neutrino sector.
Note that the total light seesaw Lagrangian (including 
the neutrino Yukawa interaction terms)
conserves $U(1)_L$, where L is Lepton number, 
if we asign the scalar singlets lepton number 2, i.e., 
$L_\varphi=L_\chi=2$ (see discussion in the next section).

The light seesaw potential in Eq.~\ref{hpotlittle} gives rise to
the desired seesaw condensate of $\Phi$
by ``coupling'' it
[through the term $\lambda_1 \left( |\Phi|^2 - |\chi|^2 \right)^2$] to 
a seesaw induced VEV of the singlet field $\chi$.
In particular, the minimization of $V(\Phi,\varphi,\chi)$ 
(assuming CP-conservation, thus setting $\xi=0$)  
which only contains terms at energy scales $\Lambda_U$ and $\Lambda_I$ 
leads to:

\begin{eqnarray}
<\varphi>  &=& \Lambda_U ~,\nonumber\\ 
<\Phi> &=& <\chi> = \frac{\Lambda_I^2}{\Lambda_U} 
\equiv v_{EW} \sim \Lambda_{EW}
\label{seesawvev1} ~,
\end{eqnarray}      

\noindent where $<\Phi>=v_{EW}$ is the condensate required for EWSB.

After EWSB (by $<\Phi>=v_{EW}$)
the $W$ and $Z$ gauge-bosons acquire their masses in the usual way 
by ``eating'' the nonphysical 
charged and neutral Goldstone fields of the SU(2) doublet. We are 
then left with three CP-even and two CP-odd physical neutral scalar states.
Expanding around the vacuum, using ({\ref{seesawvev1}) 
with $v_{EW}=<\Phi>=<\chi>$ and 
$v_\varphi=<\varphi>$, we can write:

\begin{eqnarray}
\Phi &=& \left( \begin{array}{c} 0 \\ v_{EW} + h_\Phi \end{array} 
\right) ~,\nonumber \\
\varphi &=& v_{\varphi} + h_\varphi + ia_\varphi ~,\nonumber \\
\chi &=& v_{EW} + h_\chi + ia_\chi \label{fields1}~.
\end{eqnarray}

\noindent    
Then, diagonalizing the scalar mass matrix 
we obtain the physical states which are (up to corrections 
smaller than or of ${\cal O}(\epsilon^2)$):

\begin{eqnarray}
&&{\rm CP-even}:~H \approx h_\Phi ~,~ S_1 \approx h_\chi  ~,~ S_2 \approx h_\varphi 
~,\nonumber \\
&&{\rm CP-odd}: ~A_1 \approx a_\chi ~,~ A_2 \approx a_\varphi
\label{mixH1}~,
\end{eqnarray}

\noindent with masses (up to small 
corrections of ${\cal O}(\epsilon^4)$): 

\begin{eqnarray}   
M_H &\sim& 2\sqrt{\lambda_1} \frac{\Lambda_I^2}{\Lambda_U} = 2\sqrt{\lambda_1} v_{EW} 
~, \nonumber \\
M_{S_1} &\sim& \sqrt{\lambda_3} \Lambda_U ~, \nonumber \\
M_{S_2} &\sim& 2 \sqrt{\lambda_2} \Lambda_U  ~, \nonumber \\
M_{A_1} &\sim& \sqrt{\lambda_4} \Lambda_U ~,\nonumber\\
M_{A_2} &=& 0
\label{massH1}~.
\end{eqnarray}  

\noindent Thus, the SM-like Higgs, $H$, acquires a mass of 
order $\Lambda_{EW}$ 
while the two CP-even 
physical singlets $S_1$ and $S_2$ become superheavy, i.e., with masses 
of the order of the fundamental scale $\Lambda_U$. Also, there is 
one superheavy 
axial singlet $A_1$, and a massless axial state $A_2$ which is
the Majoron associated with the
spontaneous breakdown of Lepton number if 
$L_\varphi=L_\chi=2$, see next section.
Thus, at energies of ${\cal O}(\Lambda_{EW})$, 
the light seesaw model reproduces the one-light-Higgs
SM.

\subsection{The Calmet model}

An alternative approach that falls into the category 
of seesaw Higgs models was suggested by Calmet in \cite{calmet}. 
Calmet's model is simply
a two Higgs doublet model with a seesaw Higgs mass matrix 
embedded in the scalar potential:

\begin{eqnarray}
V(\Phi_1 \Phi_2) = \pmatrix{\Phi_1^\dagger \Phi_2^\dagger} 
\pmatrix{ 0 & \Lambda_I^2 \cr \Lambda_I^2 & \Lambda_U^2 }
\pmatrix{\Phi_1 \cr \Phi_2}
+ \lambda_1 \pmatrix{\Phi_1^\dagger \Phi_1}^2
+ \lambda_2 \pmatrix{\Phi_2^\dagger \Phi_2}^2 \label{calmass}~,
\end{eqnarray}

\noindent where, 
one assumes a massless
doublet $\Phi_1$ (see footnote 1) and a superheavy 
second doublet $\Phi_2$, with a mixing interaction 
term proportional to the intermediate high scale 
$\Lambda_I$. Minimizing the Calmet's potential in (\ref{calmass}) one finds:

\begin{eqnarray}
<\Phi_1>  &\sim& \frac{1}{\sqrt{2 \lambda_1}}\frac{\Lambda_I^2}{\Lambda_U} \equiv v_{EW} 
~,\nonumber\\ 
<\Phi_2> &\sim& -\frac{1}{\sqrt{2 \lambda_1}}\frac{\Lambda_I^4}{\Lambda_U^3} 
\label{seesawvev2} ~.
\end{eqnarray} 

\noindent Thus, here also the EWSB is triggered
by the $\Lambda_U-\Lambda_I$ 
seesaw induced VEV of the light SU(2) doublet $v_{EW}= <\Phi_1>$, while
the second supermassive doublet acquires a tiny - ``double seesawed'' - VEV,  
$<\Phi_2> \sim {\cal O}(\epsilon^2 v_{EW})$, which, as will be 
shown in the next section, may be responsible for generating the sub-eV 
neutrino mass scale. 
Like any other two Higgs doublets model, after EWSB 
the physical scalar spectrum in the Calmet model consists 
of three neutral scalars (two CP-even and one CP-odd) and two 
charged Higgs states. In particular, 
one light CP-even SM-like Higgs $h$ with 
$M_h \sim 2 \sqrt{\lambda_1} v_{EW} \sim {\cal O}(\Lambda_{EW})$,
and four superheavy Higgs states $H$, $A$ and $H^\pm$
[i.e., two neutral CP-even ($H$) and CP-odd ($A$) ones and two charged ones
($H^\pm$)],  
which form the supermassive SU(2) doublet $\Phi_2$ and, therefore,  
acquire ${\cal O}(\Lambda_U)$ masses. The supermassive scalar states
decouple from the model at energy scales much smaller than the 
fundamental scale $\Lambda_U$.    
Thus, similar to our light seesaw model,  
the Calmet's model also reproduces the one-light-Higgs SM 
at the EW scale.

\section{Neutrino masses}\label{neutrino}

Introducing right handed neutrinos, the neutrino-scalar Yukawa Lagrangian 
in our light seesaw model takes the form:\footnote{Note 
that the second singlet $\chi$  
can also couple to the right handed neutrinos
via $\chi \bar{\nu_R^c} \nu_R$.
However, since $\chi$ forms a condensate of ${\cal O}(\Lambda_{EW})$, its 
contribution to the Majorana neutrino mass term will be negligible compared 
to that of $\varphi$ which forms the condensate of ${\cal O}(\Lambda_U)$.}

\begin{eqnarray}
{\cal L}_\nu = - Y_D \ell_L \Phi \nu_R + 
Y_M \varphi \bar{\nu_R^c} \nu_R \label{nulag} + h.c. ~,
\end{eqnarray}

\noindent where $\ell_L$ is the left handed SU(2) lepton doublet, 
$\nu_R$ is the right handed neutrino field, $\Phi$ and $\varphi$ are 
the SU(2) scalar doublet and singlet, respectively, and 
$Y_D$, $Y_M$ are the usual Dirac and Majorana-like Yukawa couplings.   

Thus, 
when the singlet $\varphi$ forms the 
condensate $<\varphi> = \Lambda_U$ the second term in (\ref{nulag}) will lead 
to a right handed Majorana mass which will naturally be of that 
order: $m_\nu^M=Y_M \Lambda_U$. 
The SU(2) condensate $<\Phi>=\Lambda_I^2/\Lambda_U \sim \Lambda_{EW}$ 
will generate a Dirac mass for the neutrinos 
of size $m_\nu^D \sim Y_D \Lambda_{EW}$ through the first term 
in (\ref{nulag}), .  
Then, the neutrino 
mass matrix becomes: 

\begin{eqnarray}
{\cal L}_{m_\nu} = \pmatrix{\bar{\nu_L^c} \bar{\nu_R}} 
\pmatrix{ 0 & m_\nu^D \cr m_\nu^D & m_\nu^M }
\pmatrix{\nu_L \cr \nu_R^c}  \label{numass}~,
\end{eqnarray} 

\noindent which, upon diagonalization (i.e., the classic seesaw mechanism) 
gives two Majorana neutrino physical states: 
a superheavy state $\nu_h$ with a mass 
$m_{\nu_h} \sim Y_M \Lambda_U$ and a superlight 
state $\nu_\ell$ with a mass:

\begin{eqnarray}
m_{\nu_\ell}=\frac{(m_\nu^D)^2}{m_\nu^M}
=\frac{Y_D^2}{Y_M} 
\frac{\Lambda_I^4}{\Lambda_U^3}
\sim\frac{Y_D^2}{Y_M} 
\frac{\Lambda_{EW}^2}{\Lambda_U}
\label{nuscale} ~.
\end{eqnarray}
  
\noindent The neutrino mass scale is, therefore, 
subject to a two-step seesaw 
mechanism,   
the first (in the scalar sector) generates the Dirac neutrino mass 
$m_\nu^D \sim Y_D \Lambda_{EW}$, which 
then enters in the off diagonal neutrino mass matrix to give the classic 
seesaw Majorana mass in (\ref{nuscale}) by a second $m_\nu^M - m_\nu^D$ 
seesaw in the neutrino mass matrix of (\ref{numass}).   
The presence of this extremely small scale, 
$m_{\nu_\ell} \sim {\cal O}(\Lambda_{EW}^2/\Lambda_U)$, well 
below the EW
scale, is therefore naturally explained in terms of the two ultra-high
scales $\Lambda_U$ and $\Lambda_I$.

As an example, let us suppose that $\Lambda_U \sim {\cal O}(10^{16})$ GeV.
As was shown in the previous section, this means that 
$\Lambda_I \sim {\cal O}(10^{9})$ will be needed in order to generate 
$\Lambda_{EW}={\cal O}(100)$ GeV.
Then, if $Y_D \sim Y_M \sim {\cal O}(1)$, 
it follows from
({\ref{nuscale}) that 
$m_\nu={\cal O}(10^{-3})$ eV,  
roughly in accord with current mixing
results \cite{neutrino1}.
A value of $\Lambda_U$ at the Planck scale could still be consistent 
with the double-seesaw sub-eV neutrino masses. In particular, if 
$Y_D^2 / Y_M \sim {\cal O}(10^{3})$ GeV,
then with $\Lambda_I={\cal O}(10^{10.5})$ GeV [which 
gives $\Lambda_{EW}={\cal O}(100)$ GeV when 
$\Lambda_U \sim {\cal O}(10^{19})$ GeV],  
we still obtain 
$m_\nu={\cal O}(10^{-3})$ eV. This may happen if either 
the Majorana mass is sufficiently below 
the Planck scale, i.e., $Y_M \sim {\cal O}(10^{-3})$, and the Dirac 
masses are of the order of the EW scale, i.e., $Y_D \sim {\cal O}(1)$, or
if the Majorana mass is of the order of the intermediate scale $\Lambda_I$, 
i.e., $Y_M \sim {\cal O}(10^{-8.5})$ and the Dirac masses are of 
${\cal O}(100)$ MeV  
(consistent with most light leptons and down quark masses) which
corresponds to $Y_D \sim {\cal O}(10^{-3})$. 
The latter possibility, in which 
$m_{\nu_h} \sim \Lambda_I \sim {\cal O}(10^{10.5})$, 
is particularly interesting since the existence of 
heavy Majorana neutrinos with masses in the range 
$10^9 - 10^{13}$ GeV may be useful for leptogenesis, i.e., for 
generating the observed baryon asymmetry through the lepton asymmetry 
which is triggered by the decays of these heavy 
Majorana neutrinos \cite{lepto}.

As was noted in the previous section, our light seesaw model
contains a Majoron 
which is the massless Goldstone
boson (i.e., the axial component of the singlet field $\varphi$)  
associated with the spontaneous breaking of a global $U(1)$
number \cite{plb98p265}.
In particular, in our case, the model
defined by the total Lagrangian in (\ref{totlag}) with
the scalar potential in (\ref{hpotlittle}) and with
the neutrino Yukawa terms in (\ref{nulag}),
conserves lepton number $L$ if both singlets $\varphi$ and $\chi$ 
carry lepton number 2, i.e., if $L_\varphi=L_\chi=2$. Thus, 
when $\varphi$ and $\chi$ form their condensates, 
lepton number is spontaneously broken 
and the associated Majoron is $A_2$ [see (\ref{fields1})-(\ref{massH1})].
Note that this massless Majoron
is phenomenologically 
acceptable since it
will escape detection (i.e., 
will decouple from the model) 
due to its extremely suppressed 
couplings to the matter and gauge fields \cite{plb98p265}.

As opposed to our light seesaw model, 
in the Calmet model the neutrinos will acquire 
only Dirac masses through interactions of the superheavy SU(2) doublet 
$\Phi_2$ 
with the neutrinos, i.e., ${\cal L}_{m_\nu} = Y_D \ell_L \Phi_2 \nu_R$.
In this model the ``double-seesaw'' mechanism required for generating
the sub-eV neutrino masses is operational already in the scalar sector. 
That is,     
when $\Phi_2$ forms its ``double-seesawed'' superlight condensate 
$<\Phi_2> \sim (2 \lambda_1)^{-1/2} \times \Lambda_I^4/\Lambda_U^3$
[see (\ref{seesawvev2})]   
the neutrino acquires a Dirac mass of that order:

\begin{eqnarray}
m_\nu^D = Y_D <\Phi_2> \sim \frac{Y_D}{\sqrt{2 \lambda_1}} \frac{\Lambda_I^4}{\Lambda_U^3} 
\sim \frac{Y_D}{\sqrt{2 \lambda_1}} \frac{\Lambda^2_{EW}}{\Lambda_U} \label{calnumass}~.
\end{eqnarray}

\noindent In particular, if
$Y_D\sim \lambda_1 \sim {\cal O}(1)$, then 
$m_\nu^D \sim {\cal O}(10^{-3})$ eV for $\Lambda_I \sim {\cal O}(10^9)$ GeV 
and $\Lambda_U \sim {\cal O}(10^{16})$ GeV [which also guarantees 
that 
$\Lambda_{EW} \sim \Lambda_I^2/\Lambda_U \sim {\cal O}(100)$ GeV]. However, 
if $\Lambda_U \sim {\cal O}(10^{19})$ GeV [which requires 
$\Lambda_I \sim {\cal O}(10^{10.5})$ 
GeV in order to reproduce the EW scale]
it is difficult to see in the Calmet 
model how ${\cal O}(10^{-3})$ eV neutrino masses 
can be generated, unless we assign a rather un-naturally large
value to $Y_D$, i.e., $Y_D \sim {\cal O}(10^3)$.    

Note also that, in the Calmet model, a Dirac type neutrino 
mass of the order 
of $\Lambda_{EW}$ would be generated if the light Higgs doublet $\Phi_1$ is
also coupled to the right handed neutrinos via $Y_D \ell_L \Phi_1 \nu_R$.
Therefore, in order to protect the sub-eV neutrino mass scale one has to 
assume that such $\ell_L \Phi_1 \nu_R$ interactions are absent.

\section{Seesaw Higgs mechanism from tiny extra dimensions}\label{ed}

The idea that the enormous hierarchy between the 
two seemingly disparate EW and Planck scales may result from 
the existence of compact extra spatial dimensions (CED) \cite{LED}, 
has gained 
intense interest in the past years, due to its novel approach.
In particular, according to the viewpoint taken in \cite{LED}, 
the EW scale is the only fundamental scale in nature and the 
effective large four-dimensional Planck scale (i.e., the weakness 
of gravity) is a result of the large size of the CED
(the bulk), through which gravity propagates: 

\begin{eqnarray}
R \sim \frac{1}{M_\star} \left(\frac{M_{Pl}}{M_\star}\right)^{2/n} 
\label{Rsize}~,
\end{eqnarray}

\noindent where $R$ is the typical radius of the CED,
$n$ is the number of CED, $M_{Pl}\sim 10^{19}$ GeV 
is the reduced four-dimensional Planck mass and $M_\star$ is the fundamental 
$(4+n)$ Planck scale. Putting $M_\star = \Lambda_{EW}$ in 
(\ref{Rsize}) implies that $R$ is in the sub-mm range.

In contrast to \cite{LED}, let us suppose that the ($4+n$) fundamental 
Planck scale is close to the GUT scale, i.e., 
$M_\star \sim 10^{16} ~{\rm GeV} >> \Lambda_{EW}$ (note that 
this is similar to the 
view taken in the ``warp'' extra-dimension scenario \cite{RS99}).
In this case, the typical size of the bulk CED
where gravity propagates is extremely small
$R(M_\star \sim 10^{16}~{\rm GeV})=10^{\frac{4}{n}-17}$ fm, and 
so, present and future searches for 
deviations from Newtonian gravity are clearly hopeless.     

Although such a scenario with ``tiny'' extra dimensions (TED) 
may seem phenomenology unattractive, 
it turns out that in the context of our light seesaw model, 
the seesaw induced EW scale is naturally obtained if 
indeed the CED are tiny and the fundamental  
scale is $\Lambda_U = M_\star \sim 10^{16}$ GeV.  
Thus, in what follows we will consider a specific 
mechanism, based on models with TED,
that can generate the desired intermediate scale
$\Lambda_I \sim 10^9$ GeV on our brane, which
then triggers the seesaw Higgs
mechanism
(when the fundamental scale is taken to be the GUT scale).

Let us suppose that the CED are populated with multiple 3-branes.
In this case, it was shown in \cite{multibranes} 
that the violation of flavor symmetries on these distant branes can be
carried out to our brane by "messenger" scalar fields that can propagate 
freely in the bulk between the branes . In particular,
after propagating through the bulk's CED 
(from the distant brane to our world), 
the profile of these messenger fields at all points on our wall (i.e., 
on the interference between the bulk and our brane)  
``shines'' the flavor violation which 
appears as a boundary condition on our $3$-brane.  
This mechanism may explain e.g., 
the smallness of some Yukawa couplings and the 
sub-eV
mass scale of the neutrinos \cite{multibranes,ma1,ma2}. 
In our case, 
this ``shining'' mechanism is utilized 
to generate the light seesaw Higgs 
model presented in section \ref{LHSM}. Somewhat similar to 
\cite{ma1}, we will assume that a messenger singlet scalar field ``shines''  
only the scalar sector in our brane. 

Let $\eta$ be a singlet scalar field which can propagate in the CED
and let $\eta_{{\cal P}^\prime}$ be another singlet 
localized to a distant 3-brane (${\cal P}^\prime$-brane), 
which is situated at $y^i=y^i_0$ in the CED 
($y^i$, $i=1-n$, is the coordinates of the CED such that
our brane is assumed to be localized at $y^i=0$). 
In particular, we assume that
$|y_0|=R$ ($R$ is the CED radius) which 
is the farthest point in the CED. 
Suppose now that some ``scalar-flavor'' symmetry $G_S$ 
is initially conserved everywhere and that 
both 
$\eta$ and $\eta_{{\cal P}^\prime}$ are charged
under $G_S$, with charges $S_{\eta}=-S_{\eta_{{\cal P}^\prime}}=1$.
In particular, these two singlets interact on the 
${\cal P}^\prime$-brane via:

\begin{eqnarray}
S_{{\cal P}^\prime} = \int_{{\cal P}^\prime} d^4 x^\prime 
M_\star^2 \eta_{{\cal P}^\prime}(x^\prime) \eta(x^\prime,y^i=y^i_0) 
\label{sprime}~,
\end{eqnarray} 

\noindent where $x^\prime$ is the coordinate in the 
${\cal P}^\prime$-brane (recall that $M_\star$ is the fundamental 
Planck scale).\footnote{We ignore possible 
self interaction terms of $\eta$ in the bulk.} 

Then, when $\eta_{{\cal P}^\prime}$ acquires a non-zero VEV in the 
${\cal P}^\prime$-brane, the ``scalar-flavor'' $S$ number 
is spontaneously violated 
and $<\eta_{{\cal P}^\prime}>$ will act as a point source, shining 
the $\eta$ field. The $S$ number violation will then be
communicated to our world through the shined value of $<\eta>$ 
on our wall 
\cite{multibranes}:

\begin{eqnarray}
<\eta(x,y^i=0)>=<\eta_{{\cal P}^\prime}(y^i=y^i_0)>
\times \Delta_n(R=|y_0|) \label{delta}~,
\end{eqnarray}

\noindent where $x^\mu$ ($\mu=0-3$) are the usual non-compact dimensions
of our brane and $\Delta_n$ is the Yukawa potential in the 
$n$ transverse dimensions \cite{multibranes}.
For example, for $n>2$, $m_\eta R <<1$ and 
$ <\eta_{{\cal P}^\prime}> \sim M_\star$, one obtains 
the following profile of shining on our brane \cite{multibranes,ma1,ma2}:

\begin{eqnarray}
<\eta> \sim \frac{\Gamma(\frac{n-2}{2})}{4 \pi^{\frac{n}{2}}}
\frac{M_\star}{(M_\star R)^{n-2}} \label{shine} ~,  
\end{eqnarray}

\noindent which now appears as a boundary condition for our 
4-dimensional scalar potential.

Consider now the light seesaw scalar potential in (\ref{hpotlittle}) 
and assume that the dimension-two operator $\varphi^\dagger \chi$ carries an 
$S$ number -2, i.e., $S_{\varphi^\dagger \chi}=-2$.
If $G_S$ is initially conserved everywhere, i.e., also in the bulk 
and on our brane, 
then $\eta$ interacts on our brane via the following $S$ number 
conserving term:

\begin{eqnarray}
S_{us} = \int_{us} d^4 x ~
\eta(x,y^i=0) \eta(x,y^i=0) \varphi^\dagger(x) \chi(x) +h.c. 
\label{sus}~.
\end{eqnarray}  

\noindent Thus, after propagating 
the distance between the ${\cal P}^\prime$-brane to our brane, 
the shined value of $\eta$, $<\eta> \equiv <\eta(x,y^i=0)>$, will
generate the term $\Lambda_I^2 {\rm Re}(\varphi^\dagger \chi)$ 
in (\ref{hpotlittle}), if $<\eta> = \Lambda_I$.
Using (\ref{shine}) with $M_\star \sim 10^{16}$ GeV and 
with $M_\star R \sim (M_{Pl}/M_\star)^{2/n}$ 
from (\ref{Rsize}), the desired (i.e., in order to get the seesaw induced 
EW scale)
intermediate scale, $<\eta> = \Lambda_I \sim 10^9$ GeV, is obtained 
if there are $n=7$ tiny extra transverse dimensions of size 
$R \sim M_\star^{-1} \sim 10^{-16}$ fm. 
The mass of the messenger field is bounded by
$m_\eta << R^{-1}$ and so it can be naturally of the 
same order of its shined VEV, 
i.e., $m_\eta \sim <\eta> = \Lambda_I \sim 10^9$ GeV.

\section{Summary}

We have proposed a scalar model - the ``light seesaw'' model -
in which the EW scale $\Lambda_{EW}$, is 
generated through a seesaw Higgs mechanism in the scalar sector. 
Assuming that the fundamental scale is close to the GUT 
or Planck scale $\Lambda_U \sim 10^{16}-10^{19}$ GeV, 
this model is constructed at an intermediate high scale 
$\Lambda_I \sim 10^9 - 10^{10.5}$ GeV, at which 
the SM $SU(2)_L \times U(1)_Y$ gauge symmetry is spontaneously 
broken by a SM-like Higgs condensate 
$<\Phi> = \Lambda_I^2/\Lambda_U \sim \Lambda_{EW}$. 
The intermediate scale $\Lambda_I$
is viewed as the scale of breaking 
of the unification group that underlies the physics at $\Lambda_U$. 

The model proposed is minimally constructed
in the sense that a minimal scalar sector which manifests the 
seesaw Higgs mechanism was assumed, along with
the usual SM gauge and matter fields and with the only addition of 
right handed neutrino fields to account for the recently verified 
non-zero neutrino masses.

We have shown that our light seesaw model has two main achievements:

\begin{enumerate}

\item It naturally explains
the huge gap or desert between the fundamental scale 
$\Lambda_U$ and the EW scale $\Lambda_{EW}$,
by a $\Lambda_U - \Lambda_I$ seesaw structure 
of the scalar potential that sets: 
$\Lambda_{EW} \sim \Lambda_I^2/\Lambda_{U}$. 
The hierarchy problem of the SM is, therefore, 
alleviated since the EW scale is not 
fundamental but is rather 
generated in terms of ultra-high energy phenomena.

\item It successfully predicts the existence of
sub-eV neutrino masses through a ``two-step''
seesaw mechanism; the first in the scalar sector:
$\Lambda_{EW} \sim \Lambda_I^2/\Lambda_{U}$ and the second in the 
neutrino sector:
$m_\nu \sim \Lambda_{EW}^2/\Lambda_{U} \sim \Lambda_I^4/\Lambda_{U}^3$.

\end{enumerate}

Thus, putting $\Lambda_U \sim 10^{16}$ GeV and $\Lambda_I \sim 10^9$ GeV
or $\Lambda_U \sim 10^{19}$ GeV and $\Lambda_I \sim 10^{10.5}$ GeV, 
our model naturally explains
the simultaneous existence of the 3 very disparate scales we observe 
in nature: 
$\Lambda_U \sim M_{GUT} - M_{Planck} \sim 10^{16} - 10^{19}$ GeV, 
$\Lambda_{EW} \sim M_W \sim {\cal O}(100)$ GeV and
$\Lambda_\nu \sim m_\nu \sim {\cal O}(10^{-2} - 10^{-3})$ eV, 
at the expense of introducing 
the intermediate physical scale $\Lambda_{I} \sim 10^9$ GeV.

Furthermore, we have shown that the mechanism of a seesaw induced EW
scale may naturally emanate from models with
{\it tiny} extra spatial dimensions of
size $R \sim M_\star^{-1}$, where $M_\star \sim \Lambda_U \sim 10^{16}$ GeV
is the fundamental multi-dimensional Planck scale.
In particular, the existence of numerous 
{\it tiny} extra dimensions can generate 
an intermediate scale of the right size (i.e., $\Lambda_I \sim 10^9$ GeV),
which then triggers the seesaw Higgs mechanism and
induces the many orders of magnitudes smaller EW scale.
More specifically, we found that our light seesaw scalar potential can 
be generated when a violation of some flavor symmetry at a distant brane is 
carried out to our brane by messenger bulk scalar fields (i.e., 
the ``shining'' mechanism \cite{multibranes}),
if there are 7 {\it tiny} compact extra dimensions
of size $R \sim 10^{-16}$ fm. 

While our light seesaw model successfully 
addresses the hierarchy problem of scales as well as
naturally explains the observed sub-eV mass scale of the neutrinos,  
we emphasize that this model requires a UV completion.
Although we have not discussed the details of a theory that can 
underly such seesaw Higgs models, 
we have outlined some of its salient features.
In particular, the technical aspect of naturalness\footnote{It should 
be understood, however, that since
the EW-scale is not fundamental in
the framework of seesaw Higgs models, the 
naturalness or fine tuning problem has 
nothing to do with the hierarchy of mass scales, but rather, it is 
just a technical obstacle that reflects our ignorance regarding the
underlying physics at energies above $\Lambda_I$.} requires 
that such seesaw Higgs models be embedded
into a more symmetric grand unified theory such that the stabilization 
of the ${\cal O}(\Lambda_{EW})$ mass of the light 
Higgs boson is ensured by some 
higher symmetry at the 
fundamental scale $\Lambda_U$. 
For example, a high-scale supersymmetric framework or
scale invariance properties of the GUT or
Planck scale physics (as suggested in \cite{bardeen}), 
may protect the EW Higgs mass scale from fine tuning.

Finally, we note that the experimental signatures (e.g., at the LHC)
of the light seesaw model will be similar to those of the SM and 
will, therefore, stand in contrast
to the expectations from e.g., SUSY theories, in which several scalars
with masses smaller than ${\cal O}(TeV)$ should be observed.

\begin{center}
Acknowledgment
\end{center}

We would like to thank Gad Eilam and Jose Wudka for 
very helpful discussions.
This work was supported in part by the US DOE Contract Nos.
DE-FG02-94ER40817 (ISU) and DE-AC02-98CH10886 (BNL).

\end{document}